\documentstyle[12pt]{ioplppt}

\input epsf.tex

\begin{document}

\title{Relic Gravitational Waves}

\author{R.A.~Battye\dag~ and E.P.S.~Shellard\ddag}

\address{\dag Theoretical Physics Group, Blackett Laboratory, Imperial College, Prince Consort Road, London SW7 2BZ, U.K.}

\address{\ddag Department of Applied Mathematics and Theoretical Physics, University of Cambridge, Silver Street, Cambridge CB3 9EW, U.K.}

\begin{abstract}
The next generation of gravitational wave detectors holds out the
prospect of detecting a stochastic gravitational wave background
generated
in the very early universe. In this article, we review the various 
cosmological processes which can lead to such a background, including
quantum fluctuations during inflation, bubble collisions in a first-order
phase transition and the decay of a network of cosmic strings. We conclude that
signals from strongly first-order phase transitions, possibly at the end of 
inflation, 
and networks of local cosmic strings are within the sensitivity of proposed
detectors. However, backgrounds from standard slow-roll 
inflation and the electroweak phase transition are too weak.
\end{abstract}

\section{Introduction}

An outstanding achievement of modern cosmology has been the detection of 
anisotropies in the cosmic microwave background \cite{smoot}. 
These anisotropies 
provide a snapshot of the universe about 400,000 years after the Hot Big Bang, 
just as the universe became transparent to electromagnetic radiation. Despite 
this observational triumph, as yet it has 
been insufficient to differentiate between competing paradigms for galaxy 
formation, nor has it shed much light on cosmological processes taking 
place before photon decoupling. There are, however, other types of radiation, such 
as gravitational radiation, which penetrate through this electromagnetic surface
of last scattering, travelling virtually 
unaffected since their emission, even when this occurred in the first few
fractions of a second. Of course, for
gravitons this remarkable transparency is due to their very weak interactions 
with ordinary matter which, in turn, makes them
difficult to observe. However, pioneering experiments have been proposed which 
could detect stochastic backgrounds of gravitational waves generated 
in the early universe over a range of frequencies.

Gravitational radiation detected with a particular frequency today would 
have been created at a characteristic time  (if generated 
by classical, causal mechanisms). Assuming that the radiation is emitted
at a time $t_{\rm e}$
before equal matter-radiation ($t_{\rm e}<t_{\rm eq}\sim 40,000$ years)
and that it is created with a wavelength comparable to the horizon
$\lambda(t_{\rm e})\sim t_{\rm e}$, then its frequency today is given 
by $f \sim z_{\rm eq}^{-1}(t_{\rm eq}t_{\rm e})^{-1/2}$ where the redshift 
$z_{\rm eq} \sim 2.3\times10^{4}\Omega_0 h^2$ and `little $h$' is the 
rescaled Hubble constant lying in the range $0.4\le h\le0.9$.  
The quantity that is measured by a particular experiment is the dimensionless 
amplitude
\begin{equation} 
\label{eqnone}
h_{\rm c}(f)=1.3\times 10^{-20}\sqrt{\Omega_{\rm g}(f)h^2}\left( {100{\rm Hz}
\over
f}\right)\,,\quad\Omega_{\rm g}(f)={f\over\rho_{\rm c}}{\partial\rho_{\rm g}
\over \partial f}\,,
\end{equation}
where $\Omega_{\rm g}(f)$ is the gravitational radiation density contribution
to the universe's density
per octave at a frequency $f$, and $\rho_{\rm c}$ is the critical energy density.
Clearly,  from (\ref{eqnone})
gravitational radiation with higher frequencies will be more difficult to 
detect.

There are currently four frequency bands available for studying
gravitational radiation \cite{thorne}. First, there are the microwave 
background anisotropies  created by tensor modes which may have been 
detected by the COBE-DMR experiment. In inflationary scenarios, the precise signal ratio due to 
tensor versus scalar modes is sensitively model-dependent, but it could be 
as high as 50\%.  Assuming that the entire COBE signal is due to gravitational 
waves implies a weak upper bound of $\Omega_{\rm g}<7\times 10^{-11}$ at 
a frequency of 
$3\times 10^{-17}h\,{\rm Hz}$. A second upper limit on $\Omega_{\rm g}$ 
comes from pulsars, since a stochastic background would lead to timing noise 
in measurements of the incoming periodic electromagnetic signal. A
recent limit is $\Omega_{\rm g}h^2<6\times 10^{-8}$ at 
$4\times 10^{-9}{\rm Hz}$ \cite{kaspi}, though the statistical veracity
of this result has been questioned with re-analyses of the same data
suggesting both stronger and weaker bounds
\cite{thorsett,mchugh}.
Thirdly, proposed ground based detectors will study 
frequencies around 100Hz with a maximum sensitivity of about 
$\Omega_{\rm g}h^2\approx 10^{-7}$ for the first generation of LIGO detectors 
and $\Omega_{\rm g}h^2\approx 10^{-10}$ for the next generation LIGO 
and VIRGO detectors.  Finally, the proposed space-based interferometer LISA will have
a sensitivity of $\Omega_{\rm g}h^2\approx 10^{-10}$ at about $10^{-3}{\rm Hz}$.

\section{Slow-roll inflation}

The inflationary paradigm \cite{guth,linde,AlbS} is strongly motivated because it 
resolves a number of shortcomings of the standard Hot Big Bang cosmology, 
including the horizon, flatness and monopole problems. However, possibly the 
most significant testable prediction of inflationary scenarios is a 
nearly scale-invariant spectrum of adiabatic density perturbations produced 
through quantum mechanical effects when the very early universe is dominated 
by the 
potential energy of some scalar inflaton field $\phi$. An important by-product 
of this process is the excitation of tensor modes resulting in a background 
of gravitational waves. The 
spectra for scalar and tensor modes created in a slow-roll inflationary 
model are the following, 
\begin{equation}
\label{spectrum}
{\cal P}_{\rm S}(\omega)={1\over\pi\epsilon(\phi_\omega)}\left({H(\phi_{\omega})
\over m_{\rm pl
}}\right)^2\,,\qquad{\cal P}_{\rm T}(\omega)={16\over\pi}\left({H(\phi_{\omega})
\over m_{\rm pl}}\right)^2\,,
\end{equation}
where $H(\phi)$ is the Hubble parameter during inflation, $\phi_\omega$ 
is the value of the inflaton field when the mode with comoving wavenumber 
$\omega=2\pi f$ leaves the Hubble radius $H^{-1}$, and the slow-roll parameter is
\begin{equation}
\epsilon(\phi)={m_{\rm pl}^2\over 4\pi}\left({H^{\prime}(\phi)\over H(\phi)}
\right)^{2}\,,
\end{equation}
with the prime $^\prime$ denoting differentiation with respect
$\phi$. Here, $\epsilon(\phi)$ is related to the speed at 
which $\phi$ rolls down the potential 
$V(\phi)$ (less than unity during inflation).
Once a particular mode is created, it is driven outside the Hubble radius (or 
`horizon') by the 
rapid expansion and it effectively `freezes' until it returns inside the horizon 
during the subsequent radiation and matter dominated eras. Lengthscales
corresponding to the size of the observed universe went outside the
horizon early when the inflaton field had a value denoted by
$\phi_{60}$;
after this there were about 60 e-foldings of expansion before the end
of inflation when $\epsilon (\phi_{\rm end})=1$.

For a specific inflaton potential $V(\phi)$, the dynamics of the Hubble 
parameter in the Hamilton-Jacobi formalism is given 
by \cite{lidsey}
\begin{equation}
\label{eom}
H^{\prime\,2}-{12\pi\over m_{\rm pl}^2} H^{2} = -{32\pi^2\over m_{\rm pl}^2} 
V(\phi)\,,\qquad\dot\phi = -{m_{\rm pl}^2\over 4\pi}H^{\prime}\,.
\end{equation}
The Hubble parameter monotonically 
decreases during slow-roll inflation, so from (\ref{spectrum}) the largest contribution to 
the gravitational wave spectrum is due to modes that were driven outside the 
horizon early in inflation and have just come back inside the horizon 
at the present day. These scales correspond to those probed by the COBE-DMR 
experiment and the observed anisotropies can be used with (\ref{spectrum}) 
to normalize $H_{60} \equiv H(\phi_{60})$ . Given this initial condition, 
(\ref{eom}) can be used to calculate the Hubble parameter during inflation
and hence the resulting 
spectrum of gravitational waves. Paradoxically, as we shall see, the larger the 
contribution of a model to the COBE signal, the smaller the signal will be 
in the frequency bands corresponding to LISA and LIGO/VIRGO; the Hubble 
parameter must decrease more rapidly in order to maintain the condition that 
there be 60 e-foldings before the end of inflation \cite{liddle}.

In order to illustrate this point, we shall calculate the spectrum of 
gravitational radiation produced by an inflationary model similar to 
polynomial chaotic inflation, but slightly modified to 
make analytic calculations tractable \cite{liddle}. Using a simple method 
to obtain the COBE normalization \cite{liddleb}, 
the  potential for this model is given by 
\begin{equation}
V(\phi)=6.7\times 10^{-10}m_{\rm pl}^4\left( 1+{240\over\alpha}\right)^{-1}
\left({\phi\over\phi_{60}}\right)^{\alpha}\left(1-{\alpha^2 m_{\rm pl}^2\over 
48\pi\phi^2}\right)\,,
\end{equation}
where $\phi_{60}=-(\alpha m_{\rm pl}/4\sqrt{\pi})(1+240/\alpha)^{1/2}$ and we 
require $\alpha > 1$. 
For this model, the slow-roll and Hubble parameters are given by 
\begin{equation}
\epsilon(\phi)={\alpha^2m_{\rm pl}\over 16\pi\phi^2}\,,\qquad
{H(\phi)\over m_{\rm pl}}=7.5\times 10^{-5}\left(1+{240\over\alpha}\right)^{-1/2}
\left({\phi\over\phi_{60}}\right)^{\alpha/2}\,.
\end{equation}
For $\omega>\omega_{\rm eq}$, substituting into (\ref{spectrum}) and 
appropriately redshifting yields
\begin{equation}
\label{chaotic}
\Omega_{\rm g}(\omega)h^2\approx 4.6\times 10^{-14}\left(1+{240\over\alpha}
\right)^{-1}\left[1-{4\over 240+\alpha}\log\left(\omega\over a_0H_0\right)
\right]^{\alpha/2}\,.
\end{equation}
For $\omega<\omega_{\rm eq}$, that is, scales 
which come inside the horizon later during the matter era, we have less redshifting 
and (\ref{chaotic}) must be multiplied by an additional factor 
$(\omega_{\rm eq}/\omega)^2$. 

To delineate the predictions of different slow-roll inflation models, one can 
use different values of $\alpha$; 
the obvious limiting cases being $\alpha=2$, the simplest chaotic
model, and $\alpha=80$ (say) for which almost all the COBE signal is 
due to gravitational waves (this case corresponds closely to power law inflation
with an exponential potential). 
The results of these cases are plotted in figure ~1. 
The $\alpha=2$ model has a low contribution to COBE but the spectrum for 
$\omega>\omega_{\rm eq}$ decreases slowly down to an amplitude of 
$\Omega_{\rm g}h^2\approx 10^{-16}$ at a few Hz. In contrast, the model 
with $\alpha=80$ has 
a much larger contribution to COBE, but has a very small amplitude for scales 
$\omega>\omega_{\rm eq}$; in the LISA frequency band 
this has already fallen to $\Omega_{\rm g}h^2\approx 10^{-24}$, eight
orders of magnitude smaller than for $\alpha=2$. Nonetheless, both these models 
produce signals well below the sensitivity of proposed interferometers 
by several orders of magnitude.

The largest possible COBE normalization provides a weak upper limit on the 
amplitude of gravitational waves for all scales 
$\omega>\omega_{\rm eq}$ which must be below the level 
$\Omega_{\rm g}h^2\approx 10^{-14}$. 
It may be possible to carefully construct a slow-roll potential 
which would almost 
achieve this limit in the LISA and LIGO/VIRGO frequency bands, but undoubtedly 
the potential would look rather contrived.  
In any case, it is clear that the prospects for a detectable 
slow-roll signal are bleak, so we shall have to consider 
less orthodox inflationary models. In the next section, we consider extended and hybrid inflation, but there are also more speculative superstring-inspired models which can give a larger gravitational wave background (for example see \cite{superstring}).

\begin{figure}[top]
\null\hskip 0.5in\relax
\epsfxsize 5.0in
\epsfbox{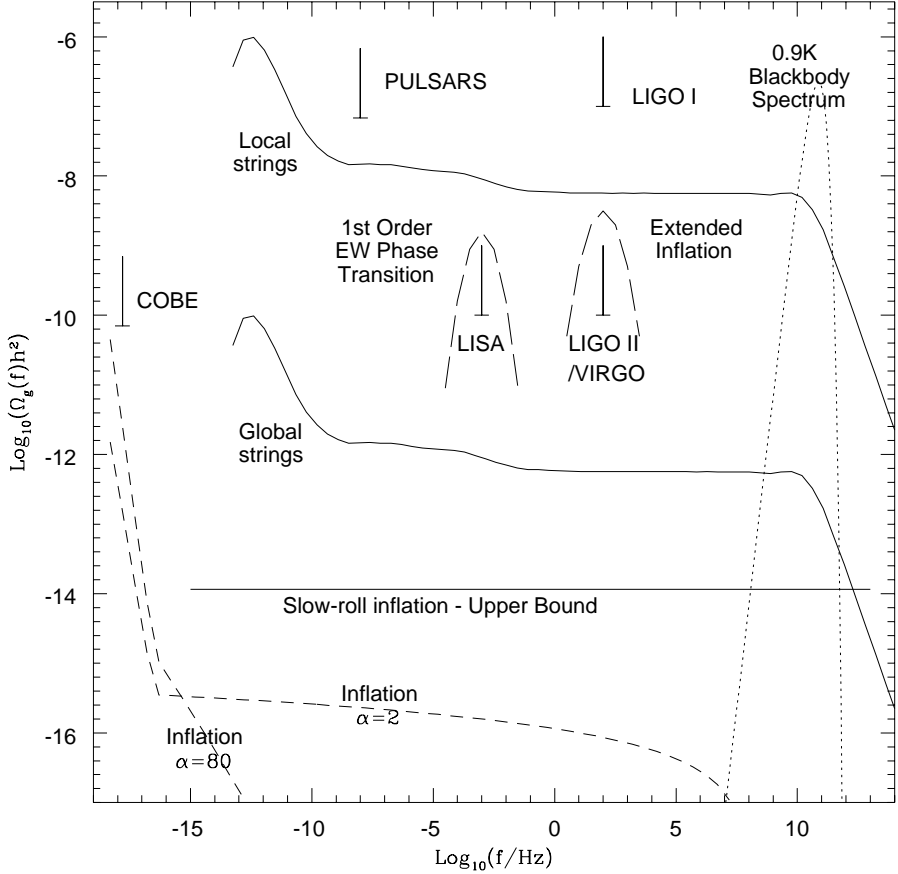}
\caption{Summary of the potential cosmological sources of a stochastic
gravitational radiation background, including inflationary models, 
first-order phase transitions and cosmic strings, as well as a primordial 0.9K 
black-body graviton spectrum (the analogue of the black-body photon radiation). 
Also plotted are the relevant constraints from the COBE measurements, 
pulsar timings, and the sensitivities of the proposed interferometers. 
Notice that local cosmic strings and strongly first-order phase transitions 
may produce detectable backgrounds, in contrast to standard slow-roll 
inflation models.}
\end{figure}

\section{First-order phase transitions}

Cosmological phase transitions in which symmetries are spontaneously broken are 
now an integral part of modern cosmology. In a first-order phase transition 
the field becomes trapped in a metastable local minimum of 
the potential---the false vacuum. The transition to the true 
vacuum takes 
place by the nucleation of bubbles and their subsequent rapid growth. 
When these vacuum bubbles collide copious amounts of gravitational 
radiation can be emitted, particularly if the relative velocity of the 
walls is relativistic as in strongly 
first-order phase transitions. It has been estimated that the maximum 
contribution to the gravitational wave background from such a transition 
would be \cite{KTW}
\begin{equation}
\Omega_{\rm g}(f_{\rm max})h^2\approx 10^{-6}\left({H_{*}\over\beta}\right)^2
\left({100\over{\cal N}_{*}}\right)^{1/3}\,,
\end{equation}
at a frequency of
\begin{equation} 
f_{\rm max}\approx 3\times 10^{-8}{\rm Hz}\left({\beta\over H_{*}}\right)
\left({{\cal N}_{*}\over 100}\right)^{1/6}\left({T_*\over 1{\rm
GeV}}\right)\,,
\end{equation}
where $\Gamma=\Gamma_0\exp(\beta t)$ is the nucleation rate of bubbles, 
$T_{*}$ is the temperature of the phase transition, $H_{*}$ is the relevant 
Hubble parameter and ${\cal N}_{*}$ is the number of relativistic degrees of 
freedom. Assuming that the electroweak phase transition is strongly first-order 
implies $\Omega_{\rm g}h^2\approx 10^{-9}$ at a frequency of 
$f_{\rm max}\approx 10^{-3}$, inside the LISA band with a detectable amplitude, 
as illustrated in figure ~1. However, the minimal standard model is 
currently believed to have
only a weakly first-order transition, if not second-order, with the
bubble walls reaching velocities well below the speed of light. 
In this case, $\Omega_{\rm g}h^2$ has been estimated to be around $10^{-22}$ 
which is well outside LISA's sensitivity \cite{KKT}.  Nevertheless, 
there are a number
of well-motivated extensions to the standard model (like supersymmetry) 
which could entail symmetry-breaking just above the electroweak scale; 
if strongly
first-order, such phase transitions would create a distinctive and detectable 
LISA signal.  Fortuitously from this point of view, LISA has a very 
interesting frequency response range.

While the prospects for detecting gravitational waves from 
slow-roll inflation seem poor, there 
are a number of inflationary models which do not end in the standard 
re-heating scenario. These include extended inflation \cite{extend} and 
hybrid inflation \cite{hybrid} which exit through a phase transition;
this provides an extra, potentially more powerful source of gravitational 
waves if 
this final phase transition is strongly first-order \cite{TW}. The radiation
could be detected by the advanced LIGO and VIRGO detectors if the re-heat 
temperature of the universe was close to $10^{8}$GeV or $10^{9}$GeV---scales 
which are far from ruled out by current observations. We have included a sample 
spectra for such a model in figure ~1.

\section{Cosmic string networks}

Cosmic strings are line-like topological defects which may form during 
phase transitions in the early universe (see, for example, 
\cite{VS}). Due to 
the large string mass per unit length (typically about $10^{22}{\rm g\,cm}^{-1}$
for GUT scale strings), they may have acted 
as the initial seeds for the formation of large-scale 
structure \cite{structureone,structuretwo}. In general, a network of strings 
will evolve towards a self-similar scaling solution by the production of loops 
and the subsequent emission of radiation into a preferred channel. For local 
strings, this channel is gravitational radiation, thus creating a 
stochastic background of radiation over frequencies from $10^{10}$Hz to 
$10^{-12}$Hz.

Scale-invariant string evolution implies that the density of 
the string network $\rho_{\infty}=\mu\zeta/t^2$ remains constant 
relative to the 
background density.  Statistically, the network 
has the same properties at any two times, except for a universal scaling 
with respect to the growing horizon size, that is, there exists a characteristic 
length scale $L=\zeta^{-1/2}t$. This scaling regime has been shown to exist 
numerically \cite{BB,AS} and various parameters have been estimated, notably
$\zeta\approx 13$ during the radiation era. 

The radiative dynamics of the strings is a crucial factor determining 
the precise contributions at various frequencies. The power radiated by a 
string loop is normally written as a sum over loop harmonics \cite{VV},
that is, 
\begin{equation}
P=\sum_{n=1}^{n_*}P_n=\Gamma G\mu^2
\end{equation}
where $\langle\Gamma\rangle\approx 65$ \cite{QSSP} is a constant parametrized 
solely by the particular loop trajectory and not its length, $\mu$ is the 
string mass per unit length and $n_{*}$ is some cut-off introduced by the 
effects of radiation backreaction \cite{BCSa} (see also \cite{BSc}).

Assuming that ${\cal N}$ is constant 
and for reasonable values of $n_*$, one can deduce that 
the spectrum of gravitational radiation due to loops created 
during the radiation era is independent of frequency and only very weakly 
dependent on the loop radiation spectrum. The amplitude of this flat 
spectrum is given by 
\begin{equation}
\Omega_{\rm g}(\omega)\approx {256\pi G\mu\over 3}\,\Omega_{\rm r}\;F\left({\alpha
\over\Gamma G\mu}\right)\,,
\end{equation}
where $F(x)=((1+x)^{3/2}-1)/x$, $\Omega_{\rm r}=\rho_{\rm r}/\rho_{\rm c}$, 
$\rho_{\rm r}=3/32\pi Gt^2$ and $\alpha$ is the constant loop production size 
with respect to the horizon. In contrast, the radiation background
today due to loops produced in the matter era is highly dependent on the 
loop radiation spectrum. Assuming that $P_{n}\propto n^{-q}$, one can 
deduce that $\Omega_{\rm g}(\omega)\propto\omega^{1-q}\rightarrow\log_{10}
[\Omega_{\rm g}(\omega)]=A+(1-q)\log_{10}[\omega]$. Therefore, the amount 
of radiation produced in the matter era which feeds into higher frequencies 
is extremely sensitive to $q$ and $n_*$. It is precisely these frequencies 
that are relevant for the pulsar timing experiment \cite{AC}. 

One can investigate these effects more fully as in ref.~\cite{BCSa} using 
a numerical algorithm which calculates the spectrum of radiation for given 
spectra and cosmological parameters \cite{AC}. The parameters that were used 
are $\Omega_0=1$, $h=0.5$ and the electroweak model particle spectrum with 
three lepton generations. It was found that the contribution to the pulsar 
timing frequency was the same for all $q\ge 2$ irrespective of the value of 
$n_*$, while for $q<2$ it was sensitive to the value of $n_*$. In particular, 
for $q=4/3$ and $n_*=\infty$ the amplitude of the spectrum will be in conflict 
with the constraint from pulsar timings for the string energy density 
$G\mu\approx 10^{-6}$ normalised 
to COBE \cite{ACSSV}. However, if $q=4/3$ and $n_*<1000$ then the contribution 
to the pulsar timing frequency will be the same as for $q=2$ and $n_*=\infty$. 
The spectrum of radiation emitted by loops is likely to be cut off by the 
effects of backreaction as shown numerically for Goldstone boson radiation 
from global cosmic strings \cite{BSc}, therefore it seems sensible to 
exclude the $q=4/3$,  large $n_*$ spectra on physical grounds (as we discuss in
ref.~\cite{BCSa}). Using well-motivated values of $q$ and $n_*$, one can use 
the pulsar 
timing constraint to deduce that $G\mu < 5.4 (\pm 0.8) \times 10^{-6}$ 
\cite{BCSa}. 
Using $\Omega_0<1$ also reduces the contribution to the pulsar timing frequency 
from matter era loops due to curvature domination at 
$t_{\rm c}\approx \Omega_0t_0$, though we expect this effect to be
small.

The contribution from cosmic strings to the LISA and LIGO/VIRGO frequency ranges 
is likely to be slightly lower than that relevant for pulsar timing because 
particle 
mass thresholds cause entropy transfers which reduce the relative density 
contribution of any pre-existing decoupled radiation. 
The cosmic string spectrum shown in 
figure ~1 illustrates this effect with a gradual rise in the otherwise flat 
spectra 
at frequencies around $10^{-4}$Hz. This particular rise is caused by 
particle annihilation near the QCD and electroweak phase transitions. If the 
particle physics model has more degrees of freedom at higher energies there 
could be other steps related to other phase transitions. However, the dependence
on ${\cal N}$ is reasonably weak and therefore we can conservatively estimate 
that  $\Omega_{\rm g}h^2>2.0\times 10^{-9}$ at $10^{-3}$Hz and $\Omega 
_{\rm g}h^2>1.0\times 10^{-9}$ at 100Hz for $G\mu = 1.0\times 10^{-6}$. 
We note also that very precise determinations of the 
stochastic background from cosmic strings at different frequencies would 
measure $\cal N$ in different cosmological epochs, providing 
fascinating insight into the particle content of the early universe (at times
as early as $10^{-25}{\rm sec}$ for LIGO and VIRGO).

Finally, we should note that there are other types of cosmic strings which do not 
produce a large background of gravitational radiation, such as those formed 
when global symmetries are broken. The dynamics of global strings is dominated by 
the production of Goldstone bosons and a simple calculation suggests that 
gravitational waves will be suppressed
by approximately 
four orders of magnitude; the global string spectrum is also illustrated in figure ~1.

\section{Conclusion}

Gravitational waves potentially provide a unique and unparallelled probe of the
very early universe.  The proposed interferometers, both terrestrial and
space-based, could rule out or severely constrain a number of viable
theoretical models, most notably the cosmic string scenario.  On the other hand,
the detection of a primordial background of gravitational waves would
have a profound impact on our understanding of high energy physics and 
cosmology, providing an 
unprecedented view of the earliest moments after the creation of the Universe.

\section*{Acknowledgements}

We would like to thank Rob Caldwell for his collaboration in reference 
\cite{BCSa} 
and we would like to acknowledge helpful discussions with 
Bruce Allen and Alex Vilenkin. 
RAB is supported by PPARC personal fellowship grant GR/K94799 and EPS is 
currently a PPARC Advanced Fellow.  This work has also been supported by
PPARC rolling grant GR/K29272.

\def\jnl#1#2#3#4#5#6{\hang{#1 [#2], {\it #4\/} {\bf #5}, #6.}}
\def\prep#1#2#3#4{\hang{#1 [#2], #4.}} 
\def\proc#1#2#3#4#5#6{{#1 [#2], in {\it #4\/}, #5, eds.\ (#6).}}
\def\book#1#2#3#4{\hang{#1 [#2], {\it #3\/} (#4).}}
\def\jnlerr#1#2#3#4#5#6#7#8{\hang{#1 [#2], {\it #4\/} {\bf #5}, #6.
{Erratum:} {\it #4\/} {\bf #7}, #8.}}
\def\prl{Phys.\ Rev.\ Lett.}
\def\pr{Phys.\ Rev.}
\def\pl{Phys.\ Lett.}
\def\np{Nucl.\ Phys.}
\def\prp{Phys.\ Rep.}
\def\rmp{Rev.\ Mod.\ Phys.}
\def\cmp{Comm.\ Math.\ Phys.}
\def\mpl{Mod.\ Phys.\ Lett.}
\def\apj{Ap.\ J.}
\def\apjl{Ap.\ J.\ Lett.}
\def\aap{Astron.\ Ap.}
\def\cqg{Class.\ Quant.\ Grav.} 
\def\grg{Gen.\ Rel.\ Grav.}
\def\mn{M.$\,$N.$\,$R.$\,$A.$\,$S.}
\def\ptp{Prog.\ Theor.\ Phys.}
\def\jetp{Sov.\ Phys.\ JETP}
\def\jetpl{JETP Lett.}
\def\jmp{J.\ Math.\ Phys.}
\def\cupress{Cambridge University Press}
\def\pup{Princeton University Press}
\def\wss{World Scientific, Singapore}

\end{document}